\documentclass[preprintnumbers]{revtex4}
\UseRawInputEncoding
\usepackage [latin1]{inputenc}
\usepackage{amssymb}
\usepackage{amsmath}
\usepackage{graphicx}
\usepackage{dcolumn}
\usepackage{bm}
\usepackage{subfigure}
\usepackage{color}
\usepackage{array}

\begin{document}

\title{Continuous Phase Transition of Higher-dimensional de-Sitter Spacetime with Non-linear Source}
\author{Yun-Zhi Du,$^{1,2}$ Huai-Fan Li,$^{1,2,*}$ Li-Chun Zhang$^{1,2}$}
\address{$^1$Department of Physics, Shanxi Datong University, Datong 037009, China\\
$^2$Institute of Theoretical Physics, Shanxi Datong University, Datong, 037009, China }

\thanks{\emph{e-mail: duyzh13@lzu.edu.cn, huaifan999@sxdtdx.edu.cn (corresponding author), zhlc2969@163.com}}

\renewcommand\arraystretch{1.5}

\begin{abstract}
For the higher-dimensional dS spacetime embedded with black holes with non-linear charges, there are two horizons with different radiation temperatures. By introducing the interplay between two horizons this system can be regarded as an ordinary thermodynamic system in the thermodynamic equilibrium described by the thermodynamic quantities ($T_{eff},~P_{eff},~V,~S,~\Phi_{eff}$). In this work, our focus is on the thermodynamic properties of phase transition for the four-dimensional dS spacetime with different values of the charge correction $\bar\phi$. We find that with the increasing of the non-linear charge correction the two horizons get closer and closer, and the correction entropy is negative which indicates the interaction between the two horizons stronger and stronger. Furthermore, the heat capacity at constant pressure, isobaric expansion coefficient, and the isothermal compression coefficient have the schottky peak at the critical point. However, the heat capacity as constant volume for the dS spacetime is nonzero. Finally, the dynamical properties of phase transition for this system have investigated based on Gibbs free energy, where exists the different behavior with that for AdS black holes.
\end{abstract}
\maketitle

\section{Introduction}

Recently people have paid more attention on the thermodynamic properties of AdS black holes and dS spacetime with black hole in order to explore the microstructure of black holes and the evolution of spacetime. By regarding the cosmological constant in a $n$-dimensional AdS spacetime as pressure ($P=-\frac{\Lambda}{8\pi}=\frac{n(n-1)}{16\pi l^2}$), which is the corresponding conjugate thermodynamic parameter of volume, the thermodynamic properties of AdS black holes were investigated \cite{Caldarelli2000,Mann1207,Cai1306,Wei1209,Wei2015,Banerjee1109,Hendi1702, Bhattacharya2017,Zeng2017,Hendi1803}. When adopting the independent parameters ($P-V$) in the charged AdS black holes, it was found that they have similar phase transition to that of van der Waals (vdW) system \cite{Mann1207,Cai1306,Wei1209,Wei2015,Banerjee1109,Zhang1502,Cheng1603,Zou1702,Hendi1702,Bhattacharya2017,Zeng2017,Hendi1803}. Along with more general vdW behaviour with standard critical exponents \cite{Dolan2014}, a broad range of other 'chemical' black hole behaviour was subsequently discovered, including reentrant phase transitions \cite{Altamirano2013}, tricritical points \cite{Altamirano2014}, Carnot cycles \cite{Johnson2014}, isolated critical points \cite{Frassino2014,Dolan2014a}, extensions to black rings \cite{Altamirano2014a}, and superfluidity \cite{Hennigar2017}.

As is well known our universe in the early period of inflation was a quasi-dS spacetime. The cosmological constant term introduced in a dS spacetime is the contribution of vacuum energy, which is also a form of material energy. When regarding the cosmological constant as dark energy, our universe will evolve into a new dS phase. In order to construct the entire history of the evolution for our universe, we should have a clear understanding of dS spacetime. However this is a tricky problem, since the absence of a killing vector that is everywhere timelike outside the black hole horizon renders a good notion of the asymptotic mass questionable. Furthermore, the presence of both the black hole horizon and cosmological horizon that of two distinct hawking temperatures suggests that the system does not meet the requirements of thermodynamic equilibrium stability. These features bring certain difficulties when we investigate the thermodynamic properties of dS spacetime. There have been a few attempts to overcome this problem and to investigate the thermodynamics of black holes in dS spacetime. One approach involves studying the horizons in a dS spacetime separately, considering them as separate thermodynamic systems characterized by their own temperature and thermodynamic behaviours \cite{Kubiznak2016,Dolan2013,Cai2002,Cai2002a,Sekiwa2006,Ali2019}. Others have investigated putting the black hole inside a cavity and imposing thermodynamic equilibrium in this closed system \cite{Carlip2003,Braden1990,Simovic2018}, which solves the thermodynamic equilibrium problem at the price of isolating the cosmological horizon and ignoring its contribution to thermodynamics of the system. In addition, authors \cite{Mbarek2019} had added scalar hair to the dS black hole to ensure equilibrium between two horizons, where only a reverse Hawking-Page transition is possible and the total entropy is the sum of two horizons (i.e., ignoring the interaction between two horizons). On this basis, people proposed that the total entropy in a dS spacetime with the thermodynamic equilibrium should contain the interaction between two horizons \cite{Guo2020,Zhang2019,Ma2020,Dinsmore2020}. From this viewpoint, we will investigate the thermodynamic phase transition for a dS spacetime with non-linear source as an ordinary thermodynamic system in the thermodynamic equilibrium.

With the development of phase transition and microstructure for various AdS black holes and the dS spacetime embedded black holes, people attempt to probe the dynamic process of AdS black hole phase transition. Recently from the view point of the Gibbs free energy, the authors in Ref. \cite{Li2020a} probed the dynamics of switching between the coexistence black hole phases by solving the Fokker-Planck equating with different reflection/aborption boundary conditions and initial conditions, and calculating the mean first passage time. In this approach, the phase transition is due to the thermal fluctuation and $G$ is regarded an function of black hole horizon which is the order parameter of phase transition. Subsequently, this method was applied to the HP phase transition in Einstein gravity \cite{Li2021} and in massive gravity \cite{Li2020} and the large/small black hole phase transition in Gauss-Bonnet gravity \cite{Wei2021}, in Einstein gravity \cite{Yang2021} minimally coupled to nonlinear electrodynamics \cite{Kumara2021}, and in dilaton gravity \cite{Mo2021}. However there is no works on the dynamic phase transition for dS spacetime embedded black holes. In this paper from the view of Gibbs free energy, we firstly attempt to exhibit the dynamic process phase transition of the four-dimensional dS spacetime embedded the black hole with non-linear charge.

Since the most physical systems are inherently non-linear in the nature, the non-linear field theories are of interest to different branches of mathematical physics. The main reason to consider the non-linear electrodynamics (NLED) is that the structures of these theories are considerably richer than the Maxwell field, and in special case they can reduce to the linear Maxwell theory (LMT). Various limitations of LMT (the self-interaction of virtual electron-positron pairs \cite{Heisenberg1936,Yajima2001,Schwinger1951} and the radiation propagation inside specific materials \cite{Lorenci2001,Lorenci2002,Novello2003,Novello2012}) motivate ones to consider NLED. The authors in \cite{Cavaglia2003} showed that NLED objects can remove both of the big bang and black hole singularities. Moreover, from astrophysical point of view, one finds that the effects of NLED become indeed quite important in superstrongly magnetized compact objects, such as pulsars and particular neutron stars (also the so-called magnetars and strange quark magnetars) \cite{Cavaglia2004}. Recently the authors in \cite{Hendi2015} presented the $n+1$-dimensional topological static black hole solutions of Einstein gravity in presence of the mentioned NLED and checked the first law of thermodynamics. Furthermore, they studied the stability of the solutions in both canonical and grand canonical ensembles and analyzed the effect of the non-linear charge correction on the thermodynamic properties of black hole. Therefore, there is naturally a question: does a dS spacetime with the non-linear charge source have the thermodynamic properties similar to a AdS black hole? In this work, we regard the higher-dimensional dS spacetime with the non-linear charge correction as an ordinary thermodynamic system by considering the interplay between two horizons and mainly investigate the properties of phase transition of the four-dimensional dS spacetime. And the effect of the non-linear charge correction on the phase transition is also analyzed.

In this paper, we briefly review the thermodynamic quantities of the higher-dimensional dS spacetime with the non-linear source in the thermodynamic equilibrium by introducing the interplay between the black hole horizon and cosmological horizon in Sec. \ref{two}. Then the phase diagrams in $P_{eff}-V$ and $G_{T_{eff}}-P_{eff}$ for the four-dimensional dS spacetime with different non-linear charge corrections are presented. We have shown the behaviors of the heat capacities nearby the critical point for this system undergoing the phase transition in the isocratic and isobaric processes as well as the curves of the isobaric expansion coefficient and the isothermal compression coefficient with the radio between two horizons for the different charge correction $\bar\phi$ in Sec. \ref{three}. From the inspect of Gibbs free energy, the dynamic process of phase transition is discribed in Sec. \ref{four}. Finally, a brief summary is given in Sec. \ref{five}.

\section{The higher-dimensional Topological black hole with the non-linear source}
\label{two}

In Refs. \cite{Zhang2019,Hendi2015,Hendi2015a,Hendi2015b,Zhao2021} the authors had shown the $n+1$ dimensional action of Einstein gravity of the non-linear source as
\begin{eqnarray}
I_G&=&-\frac{1}{16\pi}\int_M d^{n+1}x\sqrt{-g}[R-2\Lambda+L(F)]-\frac{1}{8\pi}\int_{\partial M}d^n x\sqrt{-\gamma}\Theta(\gamma),\label{LG}\\
L(F)&=&-F+\alpha F^2+o(\alpha^2),
\end{eqnarray}
where R and $\Lambda$ are the scalar curvature and the cosmological constant, respectively. And $L(F)$ is the Lagrangian of non-linear source with the Maxwell invariant $F^2=F^{\mu\nu}F_{\mu\nu}$, in which $F_{\mu\nu}=\partial_{\mu}A_{\nu}-\partial_{\nu}A_{\mu}$ is the electromagnetic field tensor and $A_{\mu}$ is the gauge potential. The non-linear charge parameter $\alpha$ is small, so the effects of the non-linear should be regarded as a perturbation. The $n+1$-dimensional topological black hole solutions are given by
\begin{equation}
\label{ds}
ds^2=-f(r)dt^2+\frac{dr^2}{f(r)}+r^2d\Omega^2_{n-1}
\end{equation}
with
\begin{equation}
f(r)=k-\frac{m}{r^{n-2}}-\frac{2\Lambda r^2}{n(n-1)}+\frac{2q^2}{(n-1)(n-2)r^{2n-4}}-\frac{4q^4\alpha}{(3n^2-7n+4)r^{4n-6}}.\label{f}
\end{equation}
Here $m$ is an integration constant which is related to the mass of the black hole ($M=\frac{V_{n-1}(n-1)m}{16\pi}$), and the last term in Eq. (\ref{f}) indicates the effect of the non-linearity. The asymptotical behavior of the solution is the AdS or dS provided $\Lambda<0$ or $\Lambda>0$, and the asymptotically flat solution is for $\Lambda=0$, and $k=1$. In the following we pay attention on the dS black hole. The horizons of the black hole and the cosmology ($r_+,~r_c$) are satisfied with the form $f(r_{+,c})=0$ for $\Lambda>0$. And the radiation temperatures at two horizon surfaces were given in Refs. \cite{Heisenberg1936,Hendi2015,Zhao2021}.

When regarding the higher-dimensional topological dS spacetime with the non-linear source as an ordinary thermodynamic system in the thermodynamic equilibrium, the thermodynamic quantities satisfy the first law
\begin{eqnarray}
dM=T_{eff}dS-P_{eff}dV+\Phi_{eff}dQ
\end{eqnarray}
whit the definition $Q\equiv\frac{q}{4\pi}V_{n-1}$. The corresponding thermodynamical volume and entropy read
\begin{eqnarray}
V=V_c-V_+=\frac{V_{n-1}r_c^n}{n}(1-x^n),~~~~~S=\frac{V_{n-1}r_c^{n-1}}{4}F(x) \label{SV}
\end{eqnarray}
with $x=\frac{r_+}{r_c}$, $V_{n-1}=\frac{2\pi^{n/2}}{\Gamma(n/2)}$, and
\begin{eqnarray}
F(x)=\frac{3n-1}{2n-1}(1-x^n)^{\frac{n-1}{n}}-\frac{n(1+x^{2n-1})+(2n-1)(1-2x^n-x^{2n-1})}{(2n-1)(1-x^n)}+1+x^{n-1}
=\bar f(x)+1+x^{n-1}.
\end{eqnarray}
The effective temperature, effective potential, effective pressure, and mass were shown as the following in Ref. \cite{Zhang2019,Zhao2021}
\begin{eqnarray}
T_{eff}&=&\frac{B(x,q)}{4\pi r_c x^{n-2}(1+x^{n+1})},~~~~~~~~~~~~~~~P_{eff}=\frac{g(x,q)}{16\pi r_c^2x^{n-2}(1+x^{n+1})},\label{Teff}\\
\Phi_{eff}&=&\frac{q(1-x^{2n-2})}{r_+^{n-2}(1-x^n)}
\left(\frac{1}{n-2}-\frac{4q^2\alpha(n-1)(1+x^{2n-2})}{[2(n^2-4)+(n-3)(n-4)]r_+^{2n-2}}\right),\label{Phieff}\\
M&=&\frac{V_{n-1}(n-1)r_c^{n-2}}{16\pi(1-x^n)}\left(k(x^{n-2}-x^n)+\frac{2q^2(1-x^{2n-2})}{(n-1)(n-2)r_c^{2n-4}x^{n-2}}
-\frac{4\bar\phi(1-x^{2n-4})}{[2(n^2-4)+(n-3)(n-4)]x^{3n-4}}\right),\label{M}
\end{eqnarray}
where $\bar{\phi}=\frac{q^4\alpha}{r_+^{4n-6}}$,
\begin{eqnarray}
B(x,q)&=&kx^{n-3}\left[(n-2-nx^2)(1-x^n)+2(n-1)x^n(1-x^2)\right]\nonumber\\
&&+\frac{2q^2x^{n-3}}{(n-1)(n-2)r_+^{2n-4}}\left[nx^n(1-x^{n-2})-(n-2)(1-x^{3n-2})\right]\nonumber\\
&&+\frac{4\bar{\phi}x^{n-3}}{3n^2-7n+4}\left[(3n-4)x^n(1-x^{4n-4})-4(n-1)x^n+nx^{4n-4}+3n-4\right],\\
g(x,q)&=&F'\left[k(n-2)x^{n-2}(1-x^2)-\frac{2q^2x^{n-2}(1-x^{2n-2})}{(n-1)r_+^{2n-4}}+\frac{4(3n-4)\bar\phi x^{n-3}(1-x^{4n-4})}{2(n^2-4)+(n-3)(n-4)}\right]\nonumber\\
&&-\frac{4(n-1)F\bar\phi x^{n-3}}{2(n^2-4)+(n-3)(n-4)}\left[\frac{nx^n(1-x^{4n-4})}{(1-x^n)}-nx^{4n-4}-(3n-4)\right]\nonumber\\
&&+\frac{2q^2(n-1)F}{(n-1)(n-2)r_c^{2n-4}x^{n-1}}\left[\frac{nx^n(1-x^{2n-2})}{(1-x^n)}-nx^{2n-2}-(n-2)\right]\nonumber\\
&&+k(n-1)Fx^{n-3}\left[\frac{nx^n(1-x^2)}{1-x^n}+n-2-nx^2\right].
\end{eqnarray}
Note that the total entropy in Eq. (\ref{SV}) is not only the sum of entropy at the black hole horizon and the cosmological horizon, and $\bar f(x)$ in the form of $F(x)$ represents the extra contribution from the correlations of the two horizons. We called $\bar{\phi}$ as the non-linear charge correction, which represents the effect of the non-linear source.

\section{Phase transition of the four-dimensional dS spacetime with the non-linear source}
\label{three}

In the following, we will mainly focus on the thermodynamical phase transition of the four-dimensional dS spacetime with the non-linear source. With Eq. (\ref{Teff}) and $n=3$, the effective temperature and the effective pressure can be rewritten as
\begin{eqnarray}
0&&=T_{eff}f_1(x)-\frac{f_2(x)}{r_+}+\frac{q^2f_3(x)}{r_+^3},\label{T}\\
0&&=P_{eff}f_4(x)r_+^4+f_5(x)r_+^2+q^2f_6(x)\label{P}
\end{eqnarray}
with
\begin{eqnarray}
f_1(x)\!&=&\!\frac{4\pi(1+x^4)}{1-x},~~f_3(x)=(1+x+x^2)(1+x^4)-2x^3,~~f_4(x)=\frac{8\pi(1+x^4)}{x(1-x)},\nonumber\\
f_2(x)\!&=&\!\frac{2\bar\phi}{5}\left[5(1+x+x^2)(1+x^8)+2x^3(1+x+x^2+x^3+x^4)\right]\nonumber\\
&&+k\left[(1-3x^2)(1+x+x^2)+4x^3(1+x)\right],\nonumber\\
f_5(x)\!&=&\!kx(1+x)F'/2-\frac{k(1+2x)F}{1+x+x^2}-\bar\phi F'(1+x)(1+x^2)(1+x^4)\nonumber\\
&&-\frac{2\bar\phi F(5+10x+15x^2+12x^3+9x^4+6x^5+3x^6)}{5(1+x+x^2)},\nonumber\\
f_6(x)\!&=&\!\frac{(1+2x+3x^2)F}{1+x+x^2}+x(1+x)(1+x^2)F'/2.\nonumber
\end{eqnarray}
The volume and entropy become
\begin{eqnarray}
V=\frac{V_2r_+^3(1-x^3)}{3x^2},~~~~~~S=\frac{V_2r_+^2F(x)}{4x^2}
\end{eqnarray}
with $V_2=\frac{2\pi^{3/2}}{\Gamma(3/2)}$.

When the system undergoing an isothermal and isobaric processes with Eqs. (\ref{T}) and (\ref{P}), the physical horizon radius $r_+$ satisfies the following expressions, respectively
\begin{eqnarray}
r_+&=&r_t=\frac{1}{2\cos(\theta+4\pi/3)}\sqrt{\frac{3q^2f_3}{f_2}},~~~
\theta=\frac{1}{3}\arccos\left(-\frac{q T_{eff}f_1}{2f_2^2}\sqrt{27f_2 f_3}\right),\label{rt}\\
r_+&=&r_p=\sqrt{\frac{-f_5+\sqrt{f_5^2-4 q^2P_{eff}f_4f_6}}{2P_{eff}f_4}}.\label{rp}
\end{eqnarray}
The critical point is determined by the following expressions
\begin{eqnarray}
\left(\frac{\partial P_{eff}}{\partial V}\right)_{T_{eff},q}=\left(\frac{\partial^2 P_{eff}}{\partial V^2}\right)_{T_{eff},q}=0.
\end{eqnarray}
\begin{table}[htbp]
\centering
\caption{Critical point with different charge correction $\bar\phi$ for $k=1,~q=1$}
\begin{tabular}{|c|c|c|c|c|c|c|}
\hline
\centering
charge corrections~~&$\bar\phi=0$~~& $\bar\phi=10^{-5}$~~&$\bar\phi=0.001$~~&$\bar\phi=0.002$~~&$\bar\phi=0.005$~~&$\bar\phi=0.01$ \\ \hline
$x_c=\frac{r_+}{r_c}$~~&0.65646~~&0.656476~~&0.65797735~~&0.659514~~&0.664132~~&0.6718755\\
\hline
$r^c_+$~~&2.6438~~&2.64382~~& 2.64723~~& 2.65037~~& 2.66029~~& 2.6783 \\ \hline
$r^c_c$~~&4.02735~~&4.02728~~& 4.02328~~&4.01867~~&4.00566~~&3.98631 \\ \hline
$V^c$~~&196.214~~&196.199~~& 195.083~~& 193.87~~& 190.359~~& 184.863 \\ \hline
$S^c$~~&68.2089~~&68.2078~~& 68.1809~~& 68.137~~& 68.0382~~& 67.9733 \\ \hline
$T_{eff}^c$~~&0.00863395~~&0.00863388~~& 0.00862669~~& 0.00861906~~& 0.00859394~~& 0.00854405 \\ \hline
$P_{eff}^c$~~&0.000583686~&0.000583694~~& 0.000584461~~& 0.000585188~~& 0.000587074~~& 0.000589151 \\ \hline
\end{tabular}\label{tcp}
\end{table}

By solving the above equations, we can obtain the thermodynamical critical quantities with different values of the charge correction $\bar\phi$ for $k=1,~q=1$ in Tab. \ref{tcp}. It is obvious from the Tab. \ref{tcp} that the phase transition will emerge with the condition $x=x_c$ and $r_+=r_+^c$ for fixed non-linear charge correction. The critical radio, radius (of the dS black hole horizon), and effective pressure are both increasing monotonically with the increase of the charge correction $\bar\phi$, while the critical radius (for the cosmological horizon), volume, effective temperature, and entropy are decreasing monotonically. That means that with the increasing of non-linear charge correction, the two horizons are getting closer and closer to each other through attraction and their interaction is becoming stronger and stronger, the correlation entropy $\bar f(x)$ between the two horizons is negative.

\begin{figure}[htp]
\subfigure[$\bar\phi=10^{-5}$]{\includegraphics[width=0.4\textwidth]{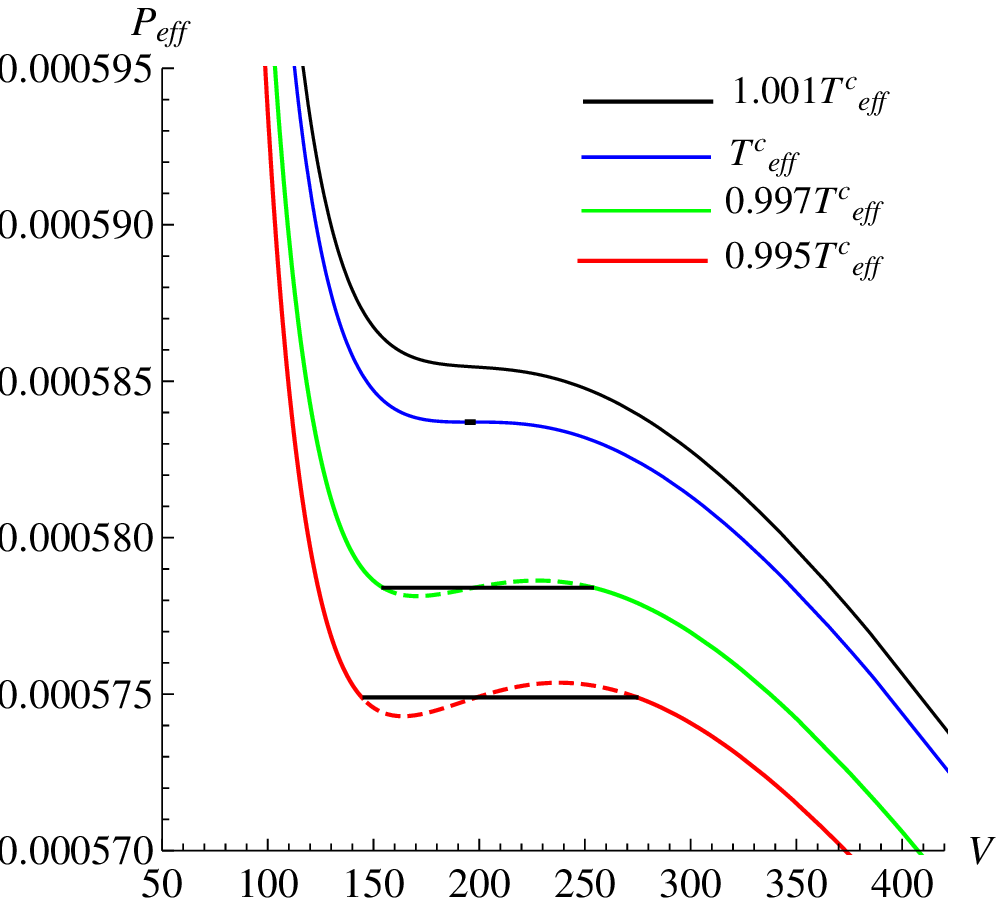}}~~~~
\subfigure[$T_{eff}=0.008591$]{\includegraphics[width=0.4\textwidth]{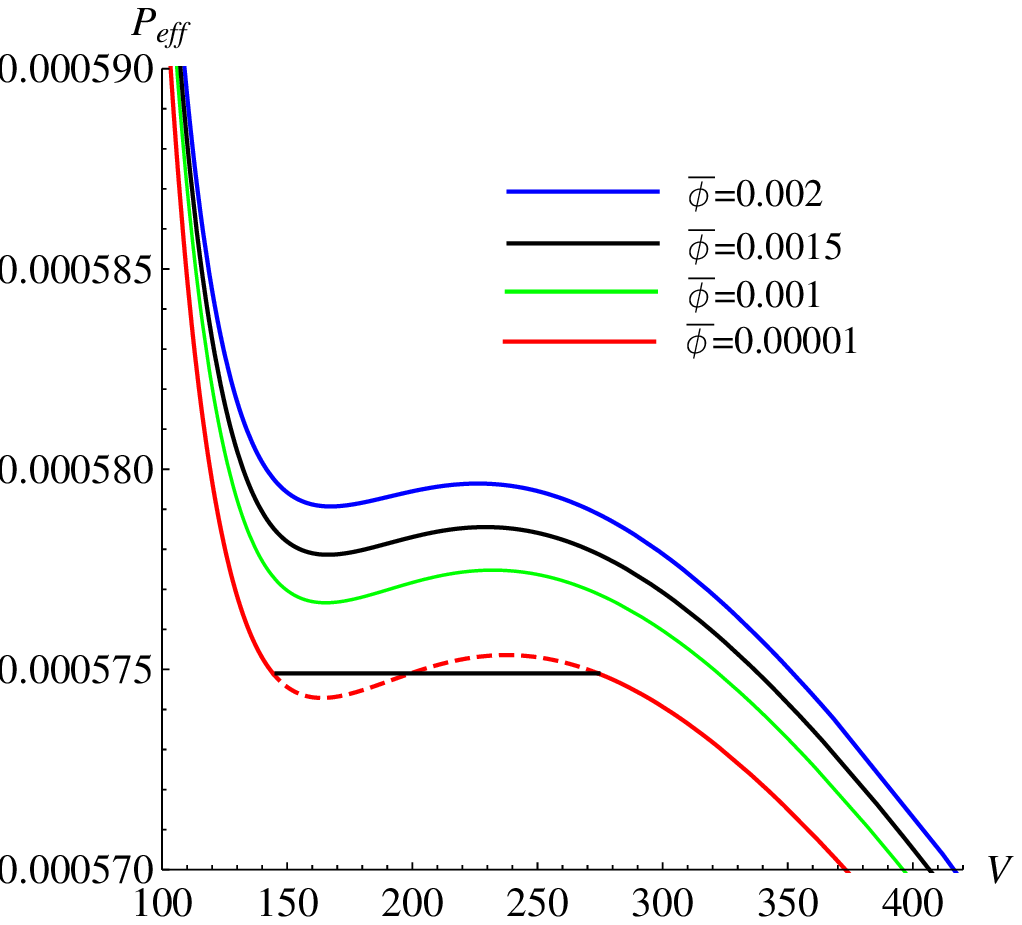}}
\caption{$P_{eff}-V$ with different non-linear corrections and different effective temperatures for the parameters $k=1,~q=1$. }\label{PV}
\end{figure}

Taking the conjugate quantities $(P_{eff}-V)$ and considering Eqs. (\ref{SV}), (\ref{T}), (\ref{P}), and (\ref{rt}), we exhibit the phase transition diagrams undergoing the isothermal processes with different non-linear charge corrections in Fig. \ref{PV}. For $T_{eff}<T_{eff}^c$,  the phase transition of the four-dimensional dS spacetime with the non-linear charge correction will emerge, while there is no phase transition for $T_{eff}>T_{eff}^c$.

Gibbs free energy is an important thermodynamic quantity to investigate the phase transition in addition to the equal area law. For the first-order phase transition, it exhibits a swallow tail behavior. At the second-order phase transition point, the Gibbs free energy is continuous but not smooth. Moreover, it also corresponds to the thermal dynamic phase transition of a black hole. From Eq. (\ref{M}) and $n=3$, we have the mass of the four-dimensional dS spacetime as
\begin{eqnarray}
M=\frac{V_2r_+(1-x^2)}{8\pi(1-x^3)}\left[k+\frac{q^2(1+x^2)}{r_+^2}-\frac{2\bar\phi}{5}\right].\label{MM}
\end{eqnarray}
And the Gibbs free energy reads
\begin{eqnarray}
G(r_+,x)=M-T_{eff}S+P_{eff}V.
\end{eqnarray}
\begin{figure}[htp]
\includegraphics[width=0.4\textwidth]{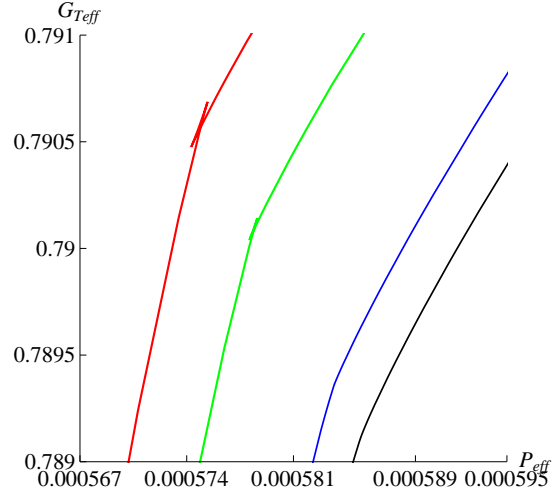}
\caption{$G_{Teff}-P_{eff}$ with the non-linear correction $\bar\phi=10^{-5}$ and $k=1,~q=1$. The effective temperature is set to $1.001T_{eff}^c$ (black lines), $T_{eff}^c$ (blue lines), $0.997T_{eff}^c$ (green lines), $0.995T_{eff}^c$ (red lines), respectively.\label{GP}}
\end{figure}
For this system undergoing the isothermal processes, considering Eqs. (\ref{SV}), (\ref{T}), (\ref{P}), (\ref{rt}), and (\ref{MM}), we present the Gibbs free energy nearby the critical points with different non-linear charge corrections in Fig. \ref{GP}. For low temperature, there appears the swallow tail behavior. At the effective critical temperature, this characterized behavior disappears. When beyond the critical point, Gibbs free energy is an increasing function of the pressure, and no phase transition emerges. Reading out the self-cross point, one can obtain the phase transition effective temperature and pressure.

\begin{figure}[htp]
\subfigure[$C_V-T_{eff}$]{\includegraphics[width=0.4\textwidth]{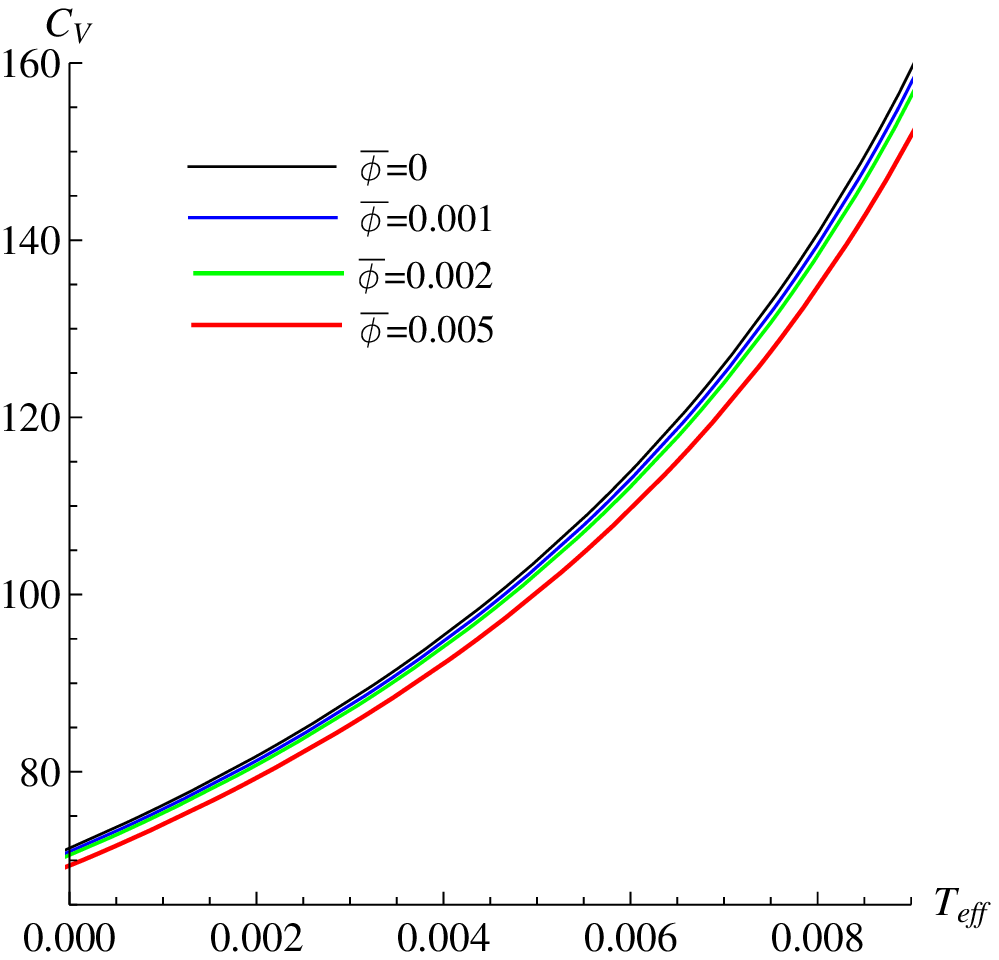}}~~~~
\subfigure[$C_{P_{eff}}-T_{eff}$]{\includegraphics[width=0.4\textwidth]{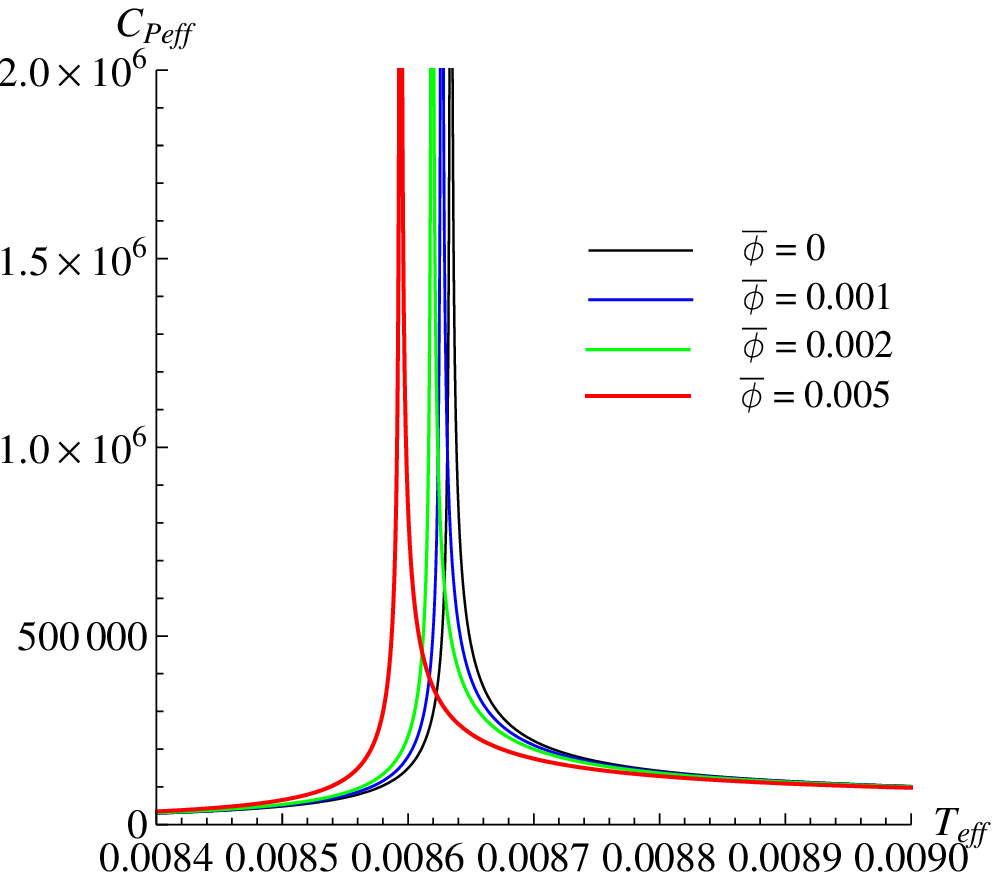}}
\caption{The heat capacities as the functions of the effective temperature $T_{eff}$ nearby the critical point with different non-linear correction for the parameters $k=1,~q=1$.}\label{CT}
\end{figure}

\begin{figure}[htp]
\subfigure[$\alpha_{P_{eff}}-x$]{\includegraphics[width=0.45\textwidth]{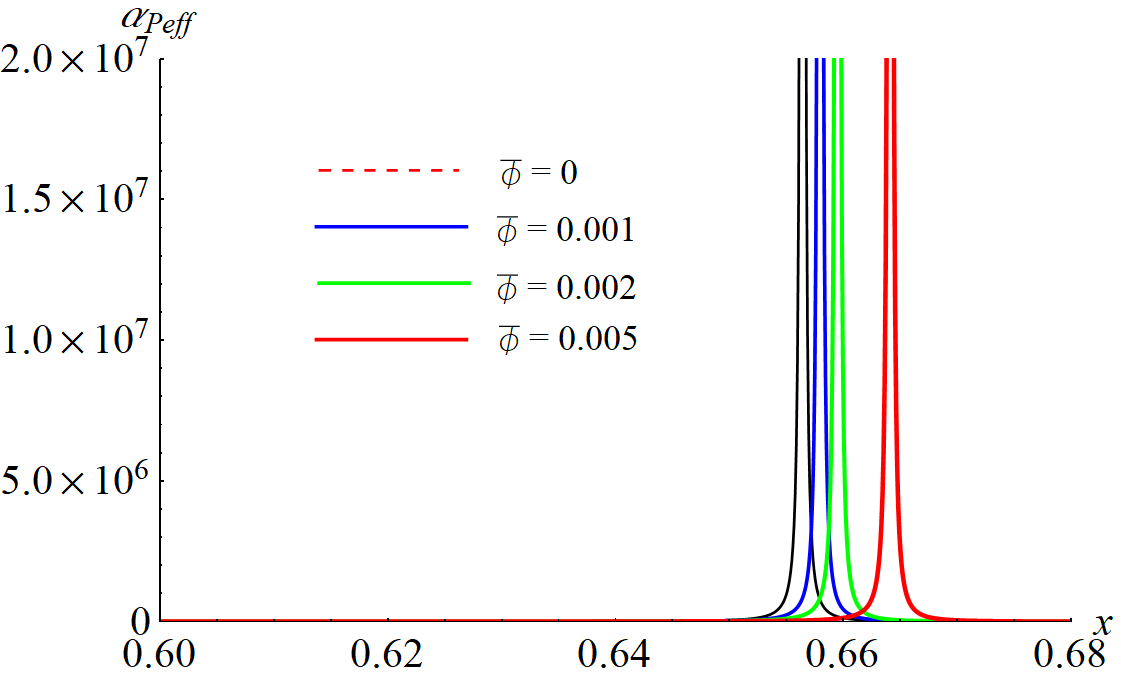}}~~~~
\subfigure[$\kappa_{T_{eff}}-x$]{\includegraphics[width=0.45\textwidth]{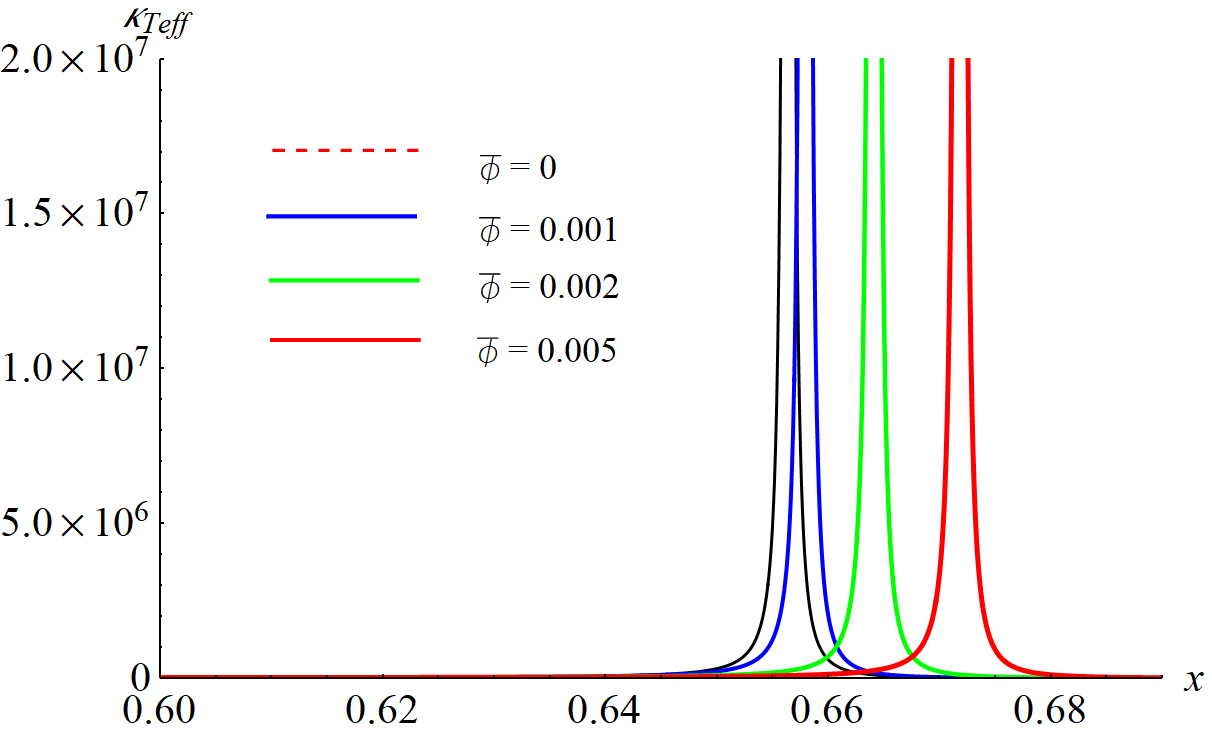}}
\caption{The isobaric expansion coefficient and the isothermal compression coefficient as the functions of the radio $x$ ($x=\frac{r_+}{r_c}$) nearby the critical point with the different charge corrections and the parameters $k=1,~q=1$.}\label{alpha-beta}
\end{figure}

The heat capacities at constant pressure and at constant volume are respectively
\begin{eqnarray}
C_V&=&T_{eff}\left(\frac{\partial S}{\partial T_{eff}}\right)_V
%=T_{eff}\frac{\frac{\partial(S,V)}{\partial(r_+,x)}}{\frac{\partial (T_{eff},V)}{\partial(r_+,x)}}
=T_{eff}\left(\frac{\frac{\partial S}{\partial r_+}\frac{\partial V}{\partial x}-\frac{\partial S}{\partial x}\frac{\partial V}{\partial r_+}}{\frac{\partial T_{eff}}{\partial r_+}\frac{\partial V}{\partial x}-\frac{\partial T_{eff}}{\partial x}\frac{\partial V}{\partial r_+ }}\right),\\
C_{P_{eff}}&=&T_{eff}\left(\frac{\partial S}{\partial T_{eff}}\right)_{P_{eff}}
%=T_{eff}\frac{\frac{\partial(S,{P_{eff}})}{\partial(r_+,x)}}{\frac{\partial (T_{eff},{P_{eff}})}{\partial(r_+,x)}}
=T_{eff}\left(\frac{\frac{\partial S}{\partial r_+}\frac{\partial {P_{eff}}}{\partial x}-\frac{\partial S}{\partial x}\frac{\partial {P_{eff}}}{\partial r_+}}{\frac{\partial T_{eff}}{\partial r_+}\frac{\partial {P_{eff}}}{\partial x}-\frac{\partial T_{eff}}{\partial x}\frac{\partial {P_{eff}}}{\partial r_+ }}\right).
\end{eqnarray}
The isobaric expansion coefficient and the isothermal compression coefficient have the following forms
\begin{eqnarray}
\alpha_{P_{eff}}&=&\frac{1}{V}\left(\frac{\partial V}{\partial T_{eff}}\right)_{P_{eff}}
%=\frac{1}{V}\left(\frac{\frac{\partial(V,P_{eff})}{\partial(r_+,x)}}{\frac{\partial (T_{eff},P_{eff})}{\partial(r_+,x)}}\right)
=\frac{1}{V}\left(\frac{\frac{\partial V}{\partial r_+}\frac{\partial P_{eff}}{\partial x}-\frac{\partial V}{\partial x}\frac{\partial P_{eff}}{\partial r_+}}{\frac{\partial T_{eff}}{\partial r_+}\frac{\partial P_{eff}}{\partial x}-\frac{\partial T_{eff}}{\partial x}\frac{\partial P_{eff} }{\partial r_+ }}\right),\\
%\beta_V&=&\frac{1}{P_{eff}}\left(\frac{\partial P_{eff}}{\partial T_{eff}}\right)_V
%=\frac{1}{P_{eff}}\left(\frac{\frac{\partial (P_{eff},V)}{(r_+,x)}}{\frac{\partial (T_{eff},V)}{\partial(r_+,x)}}\right)
%=\frac{1}{P_{eff}}\left(\frac{\frac{\partial P_{eff}}{r_+}\frac{\partial V}{x}-\frac{\partial P_{eff}}{x}\frac{\partial V}{r_+}}{\frac{\partial T_{eff}}{r_+}\frac{\partial V}{x}-\frac{\partial T_{eff}}{x}\frac{\partial V}{r_+}}\right),\\
\kappa_{T_{eff}}&=&-\frac{1}{V}\left(\frac{\partial V}{\partial P_{eff}}\right)_{T_{eff}}
%=-\frac{1}{V}\left(\frac{\frac{\partial(V,T_{eff})}{\partial(r_+,x)}}{\frac{\partial (P_{eff},T_{eff})}{\partial(r_+,x)}}\right)
=-\frac{1}{V}\left(\frac{\frac{\partial V}{\partial r_+}\frac{\partial T_{eff}}{\partial x}-\frac{\partial V}{\partial x}\frac{\partial T_{eff}}{\partial r_+}}{\frac{\partial P_{eff}}{\partial r_+}\frac{\partial T_{eff}}{\partial x}-\frac{\partial P_{eff}}{\partial x}\frac{\partial T_{eff} }{\partial r_+ }}\right).
\end{eqnarray}
With Eqs. (\ref{SV}), (\ref{T}), (\ref{P}), (\ref{rt}), and (\ref{rp}), we have exhibited the pictures of $C_V-T_{eff},~C_{P_{eff}}-T_{eff},~\alpha_{P_{eff}}-x,$ and $\kappa_{T_{eff}}-x$ in Figs. \ref{CT} and \ref{alpha-beta}.

From the Figs. \ref{CT} and \ref{alpha-beta}, for the dS spacetime with the certain non-linear charge correction $\bar\phi$ the heat capacity at constant pressure, the isobaric expansion coefficient, and the isothermal compression coefficient at the critical point are all of the schottky peak. That is consistent with the conclusions for an ordinary thermodynamic system in the thermodynamic equilibrium. Furthermore, the schottky peak in $C_{P_{eff}}-T_{eff}$ moves toward the left with the increasing of $\bar\phi$, i.e., the effective critical temperature becomes smaller with the nonlinear charge correction. In diagrams $\alpha_{P_{eff}}-x$ and $\kappa_{T_{eff}}-x$, it moves toward the right, which indicates the critical radio $x_c~(x_c=\frac{r_+^c}{r_c^c})$ becomes bigger. Those properties are fully consistent with what we had shown in Tab. \ref{tcp}. Note that it is very interesting that the heat capacity at constant volume is nonzero, which is completely contrary to that in a AdS black hole. That can be regarded a difference between dS spacetime and AdS black hole. In addition, the heat capacity at constant volume decreases with the increasing of the non-linear charge correction.

\section{Dynamic Properties of Thermodynamic Phase Transition}
\label{four}
Recently, authors in Refs. \cite{Li2020a} proposed that Gibbs free energy is related with the thermal dynamic phase transition of RN-AdS black hole. Subsequently this idea was applied to various AdS black holes \cite{Li2020,Li2021,Wei2021,Yang2021,Kumara2021,Mo2021}. Based on this, in the following we will investigate the thermal dynamic phase transition for this system from the view of Gibs free energy.

\begin{figure}[htp]
\includegraphics[width=0.4\textwidth]{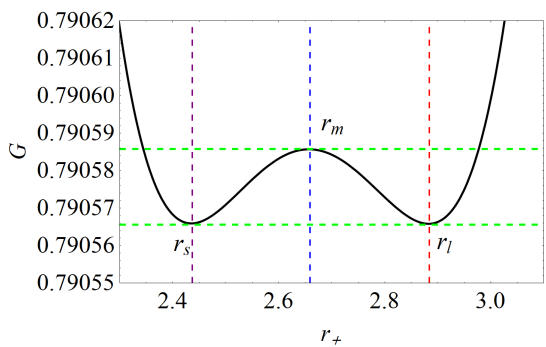}~~~~
\caption{Gibbs free energy as the function of $r_+$ with the non-linear charge correction $\bar\phi=10^{-5}$ and the parameters $k=1,~q=1$.}\label{Gr00001}
\end{figure}

The picture of Gibbs free energy at phase transition point of $P_0=0.0005749$ and $T_0=0.995Tc$ is exhibited in Fig. \ref{Gr00001}. From this picture, we can see that the Gibbs free energy displays the double-well behavior. Namely there are two local minimum (located at $r_s=2.447,~r_l=2.884$) which are corresponding to the stable small/large dS black hole with positive heat capacity. The local maximum located at $r_m=2.659$ stands for the unstable intermediate-dS black hole state with negative heat capacity and acts as a barrier between the stable small and large dS black holes. Furthermore, the depths of two local minimum are the same. It indicates that the phase transition will occur at the case of the same depth for two wells from the view of Gibbs free energy. At this issue, we expect that the reentrant phase transition or triple point maybe correspond to more wells of Gibbs free energy.

\subsection{Fokker-Planck Equation and Probabilistic Evolution}
\label{4.1}

Recently authors proposed the stochastic dynamic process of AdS black hole phase transition can be studied by the associated probabilistic Fokker-Planch equation on Gibbs free energy landscape \cite{Li2020a,Li2020,Li2021,Wei2021,Yang2021,Kumara2021,Mo2021}, which is an equation of motion governing the distribution function of fluctuating macroscopic variables. For a AdS black hole thermodynamic system, the horizon $r_+$ is the order parameter and it can be regarded as a stochastic fluctuating variable during phase transition. Based on this, by regarding the dS black hole horizon as the order parameter and a stochastic fluctuating variable for this dS spacetime we will exhibit the dynamical process of phase transition in the canonical ensemble under the thermal fluctuations. Note that the canonical ensemble is consisted of a series of dS spacetime embedded with arbitrary  dS black hole horizons.

The probability distribution of these dS spacetimes $\rho(t,r_+)$ satisfies the Fokker-Planck equation on Gibbs free energy
\begin{eqnarray}
\frac{\partial\rho(t,r_+)}{\partial t}=D\frac{\partial}{\partial r_+}\left(e^{-\beta G(r_+,x)}\frac{\partial}{\partial r_+}\left[e^{\beta G(r_+,x)}\rho(t,r_+)\right]\right), \label{rho}
\end{eqnarray}
where $\beta=1/kT$, $D=kT/\xi$ is the diffusion coefficient with $k$ being the Boltzman constant and $\xi$ being dissipation coefficient. Without loss of generality, we set $k=\xi=1$. Note that although Gibbs free energy is a function of $r_+$ and $x$, the probability distribution is still selected as a function of $r_+$ and $t$ by substituting the relationship between $x$ and $r_+$ in certain way. In order to solve the above equation, two types of boundary ($r_+=r_{0}$) condition should be imposed. One is the reflection boundary condition, which preserves the normalization of the probability distribution. The other is the absorption boundary condition.

In this system the location of left boundaries should be smaller than $r_s$, and the right one is bigger than $r_l$. Since the effective temperature of the system with the effective pressure $P_0=0.0005749$ has a minimum value $0.0081799$, there exists the minimum value of $r_+$: $r_{min}=1.90074$. We can regard $r_{min}$ as the left boundary $r_{lb}$, and set the right one $r_{rb}$ as $4$. The reflection boundary condition means the probability current vanishes at the left and right boundaries:
\begin{eqnarray}
j(t,r_0)=-T_Ee^{-G_L/T_E}\frac{\partial}{\partial r_+}\left(e^{G_L/T_E}\rho(t,r_+)\right)\mid_{r_+=r_0}=0
\end{eqnarray}
And the absorption one means the vanishing probability distribution function at the boundary: $\rho(t,r_0)=0$. The adoption of boundary condition is determined by the considering physical problem.

The initial condition is chosen as a Gaussian wave packet located at $r_i$:
\begin{eqnarray}
\rho(0,r_+)=\frac{1}{\sqrt{\pi}}e^{-\frac{(r_+-r_i)^2}{a^2}}.
\end{eqnarray}
Here $a$ is a constant which determines the initial width of Gaussian wave packet and it does not influence the final result. Since we mainly consider the thermal dynamic phase transition, $r_i$ can be set to $r_s$ or $r_l$. It means this thermal system is initially at the dS spacetime embedded with the small or large dS black hole.

\begin{figure}[htp]
\center{
\subfigure[$r_i=r_l$]{\includegraphics[width=0.55\textwidth]{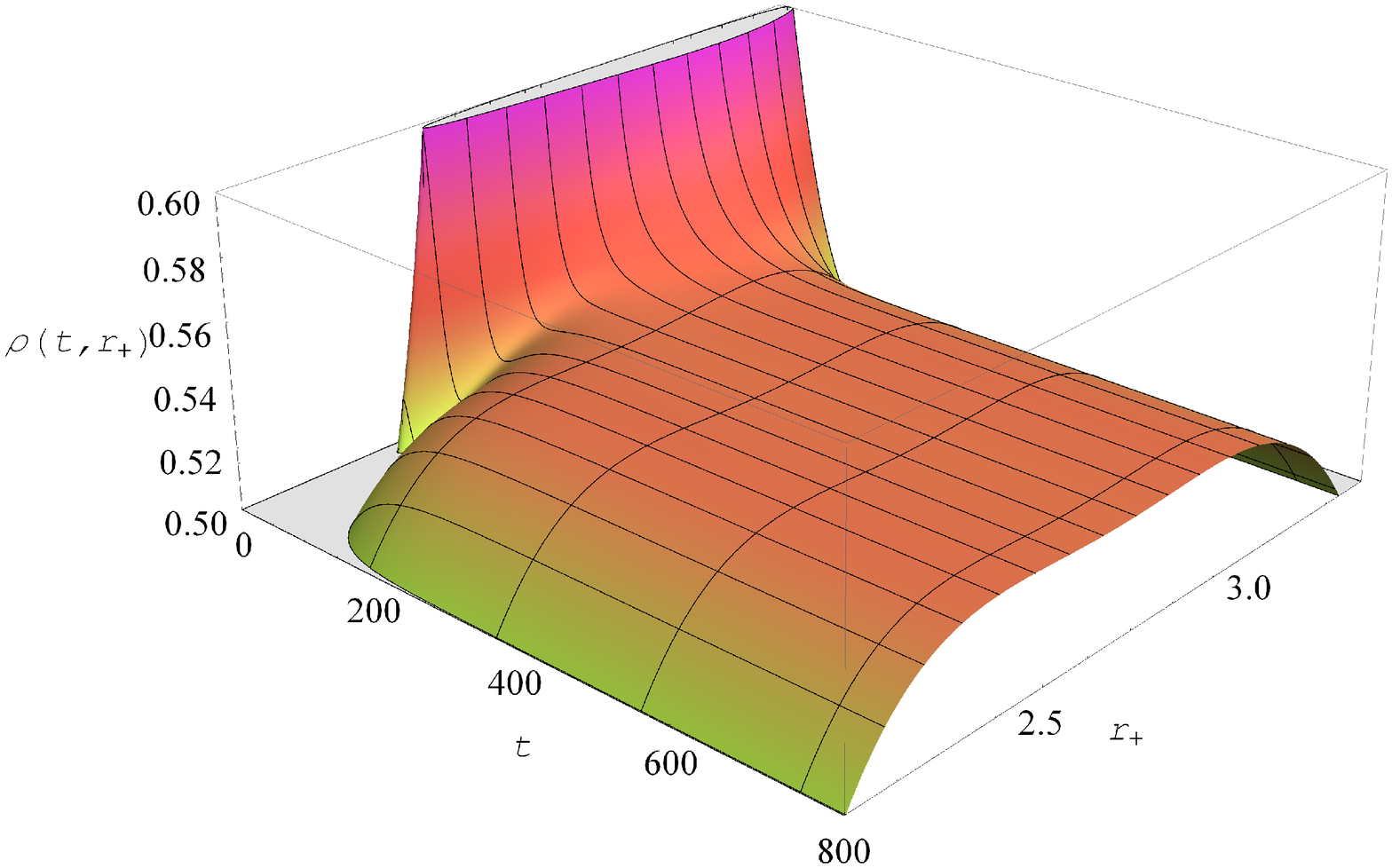}\label{rhortL00001}}
\subfigure[$r_i=r_l$]{\includegraphics[width=0.4\textwidth]{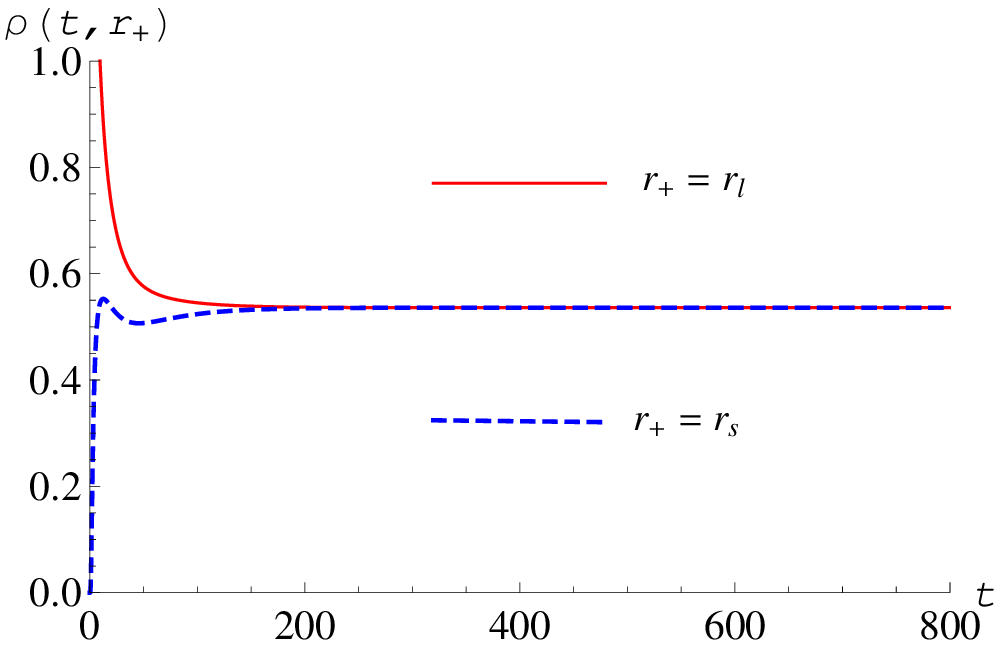}\label{rhotL00001}}}
\caption{$\rho-t-r_+$ with the parameters $k=1,~q=1,~\bar\phi=10^{-5}$. The effective temperature is set to $0.995T_c$.}\label{rhoL00001}
\end{figure}

\begin{figure}[htp]
\center{
\subfigure[$r_i=r_s$]{\includegraphics[width=0.4\textwidth]{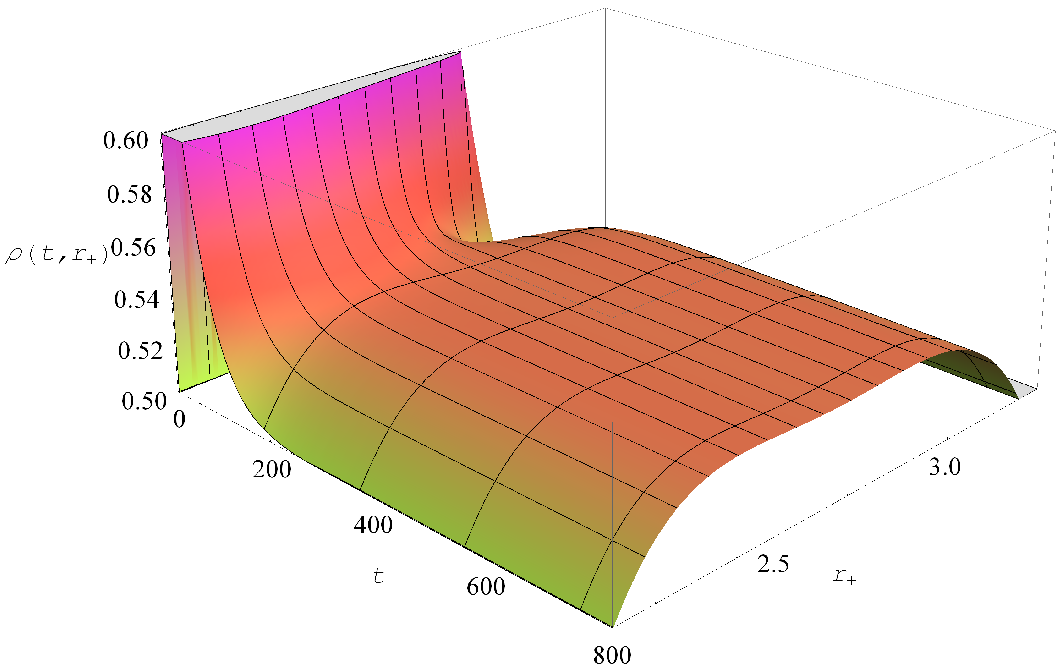}\label{rhortS00001}}~~~~~~~~
\subfigure[$r_i=r_s$]{\includegraphics[width=0.4\textwidth]{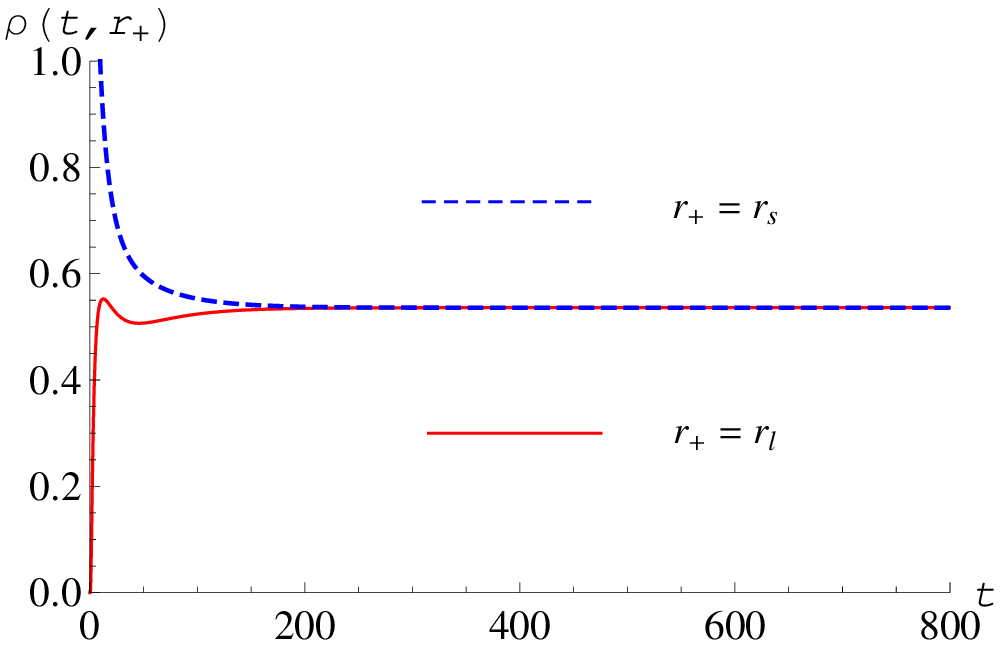}\label{rhotS00001}}}
\caption{$\rho-t$ with the parameters $k=1,~q=1,~\bar\phi=10^{-5}$. The effective temperature is set to $0.995T_c$.}\label{rhoS00001}
\end{figure}

The time evolution of the probability distribution are shown in Figs. \ref{rhoL00001} and \ref{rhoS00001}. As $t=0$ the Gaussian wave packets locate at the dS spacetime state with the large dS black hole and with the small dS black hole when $T_{eff}=0.995T_c$ and $a=0.1$, respectively. And they are both decreasing with increasing time $t$ until tending to a certain constant. However, at the same time the peaks of $\rho(t,r_+)$ at $r_+=r_{s}$ (Fig. \ref{rhotL00001}) and $r_+=r_{l}$ (Fig. \ref{rhotS00001}) are increasing from zero to the same constant. This indicates that the system embedded with large dS black hole tends to the phase with small dS black hole as shown Fig. \ref{rhotL00001}, while it embedded with small dS black hole is tending to the one with large dS black hole as shown Fig. \ref{rhotS00001}. It should be noticed that the peak behaviors of $\rho(t,r_+)$ at $r_+=r_{s}$ (Fig. \ref{rhotL00001}) and $r_+=r_{l}$ (Fig. \ref{rhotS00001}) are not monotonous increasing in a short time near the initial time comparing with a AdS black hole, which will lead to a conjecture that the microstructure of dS spacetime embedded black holes may be different from that in AdS black holes. Finally the system reaches a coexistence stationary state at a short time. This is consistent with what has shown in $G-r_+$, i.e., Gibbs free energy of the dS spacetime with large black hole and small black hole have the same depth in the double well.

\subsection{First Passage Time}
\label{4.2}
In general, the important quantity in the dynamical process of phase transition is characterized by the first passage time, which is defined as the mean value of the first passage time that the dS spacetime with a stable large black hole or a stable small black hole scape to the one with a unstable intermediate black hole (i.e., from the one well to the barrier of Gibbs free energy).

Supposing there is a perfect absorber in a stable state, if the system makes the first passage under the thermal fluctuation, the system will leave this state. We can define $\Sigma$ to be the sum probability of the dynamical process within the first passage time as
\begin{eqnarray}
\Sigma=\int^{r_m}_{r_{min}}\rho(t,r_+)dr_+, ~~~~~~or~~~~~\Sigma=-\int^{r_m}_{r_{rb}}\rho(t,r_+)dr_+,\label{sigma}
\end{eqnarray}
where $r_m$, $r_{min}$, $r_{rb}$ are the intermediate, minimum, and right boundary of charge dS black hole horizons. At a long time, the probability of this system becomes zero, i.e., $\Sigma(t,r_l)\mid_{t\rightarrow\infty}=0$ or $\Sigma(t,r_s)\mid_{t\rightarrow\infty}=0$. As claimed, the first passage time is a random variable because the dynamical process of phase transition is caused by thermal fluctuation. Hence, we denote the distribution of the first passage time by $F_p$, which reads
\begin{eqnarray}
F_p=-\frac{d\Sigma}{dt}.\label{Ft}
\end{eqnarray}
It is obviously that $F_pdt$ indicates the probability of the dS spacetime with the stable large or small black hole passing through the one with unstable intermediate black hole for the first passage time in the time interval ($t,~t+dt$). Considering Eqs. (\ref{rho}) and (\ref{sigma}), the distribution of the first passage time $F_p$ becomes \cite{Li2020a}
\begin{eqnarray}
F_p=-D\frac{\partial\rho(t,r_+)}{\partial r}\mid_{r_m}, ~~~~~or~~~~F_p=D\frac{\partial\rho(t,r_+)}{\partial r}\mid_{r_m}
\end{eqnarray}
Here the absorbing and reflecting boundary conditions of the Fokker-Planck equation are imposed at $r_m$ and the other end ($r_{min}$ or $r_{rb}$). Note that the normalisation of the probability distribution is not preserved.

\begin{figure}[htp]
\center{
\subfigure[$r_i=r_l$]{\includegraphics[width=0.4\textwidth]{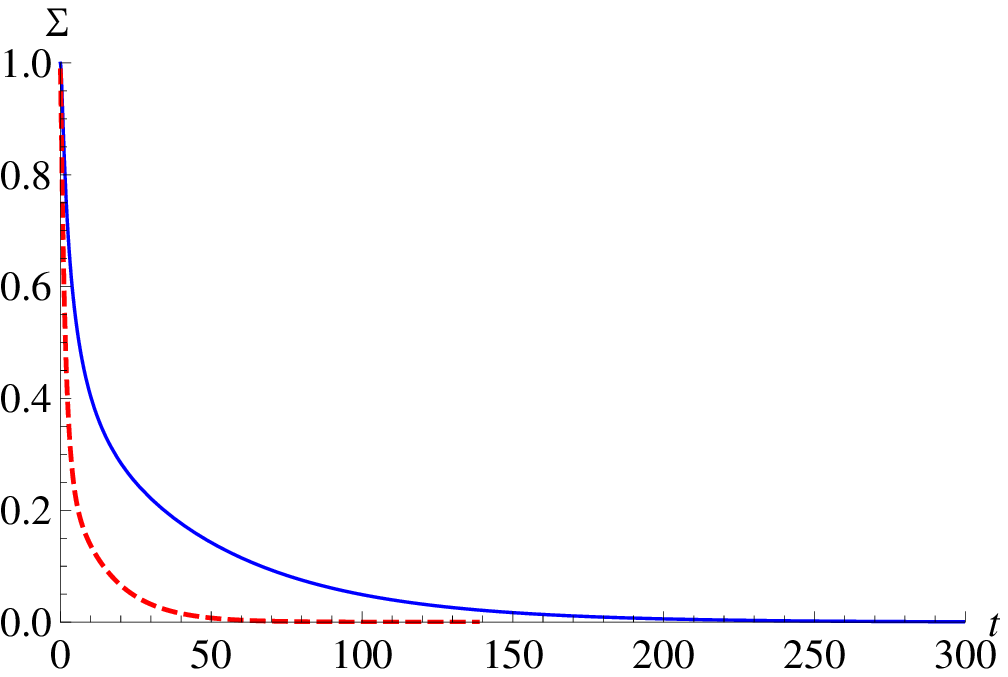}\label{sigma0}}~~~~~~~~
\subfigure[$r_i=r_l$]{\includegraphics[width=0.4\textwidth]{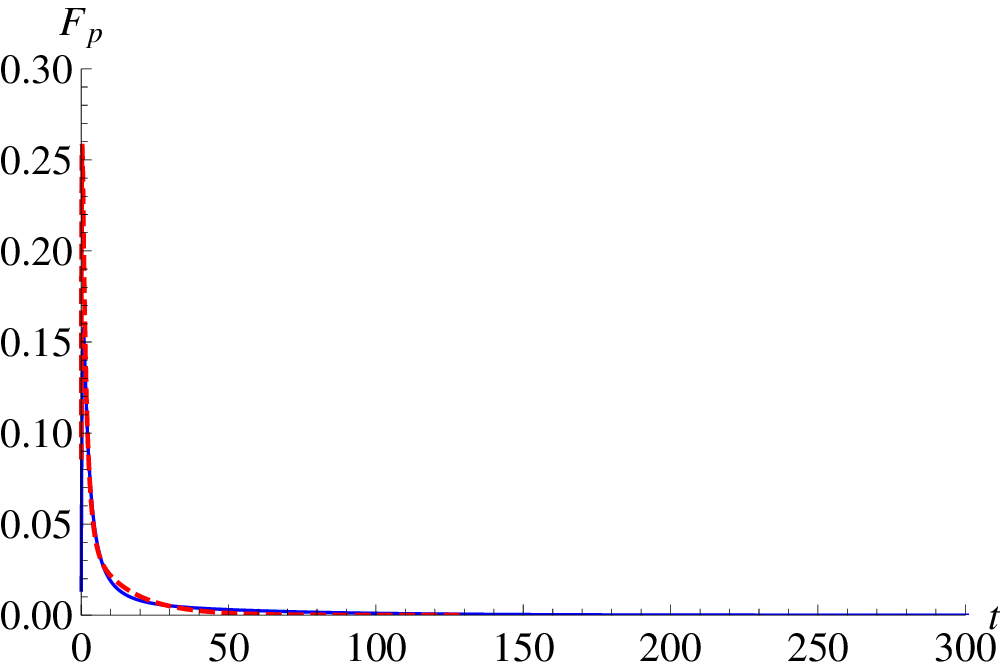}\label{Ft0}}}
\caption{The pictures of $\Sigma-t$ and $F_p-t$ with the parameters $k=1,~q=1,~\bar\phi=10^{-5}$. The effective temperature is set to $0.995T_c$ (blue lines) and $0.997T_c$ (dashed red lines), respectively.}\label{sigmaFt0}
\end{figure}

By solving the Fokker-Planck equation (\ref{rho}) with different phase transition temperatures ($T_{eff}=0.995T_c$ and $T_{eff}=0.997T_c$) and substituting them into Eqs. (\ref{sigma}) and (\ref{Ft}), the numerical results are displayed in Fig \ref{sigmaFt0} for the initial condition $r_i=r_l$. It is obviously clear that in Fig. \ref{sigma0}, $\Sigma$ decays very fast in a short time. Furthermore further increasing the effective temperature the probability of $\Sigma$ drop faster. An important point to pay is that the probability is not conserved. From the corresponding probability distribution picture in Fig. \ref{Ft0}, the behavior of $F_p$ is the similar for different effective temperatures. A single peak emerges near $t=0$ in the curve of $F_p$ for the given effective temperature. This can be understood as a large number of first passage events occur in a short interval of time, and the probability distribution decays exponentially with time. The effect of temperature at phase transition points on $F_p$ is consistent with that of $\Sigma$ and $G$. That means the higher effective temperature, the faster probability decreases, the easier phase transition occurs, and the lower depth of the barrier is; otherwise the lower temperature, the slower probability decreases, the harder phase transition occurs, and the higher depth of the barrier is.

\section{discussions and conclusions}
\label{five}

In this paper, we mainly have focused on the extended thermodynamic phase transition of the four-dimensional topological dS spacetime with the non-linear charge correction, which can be regarded as an ordinary thermodynamic system in the thermodynamic equilibrium.

Firstly we reviewed the first law of thermodynamics and effective thermodynamic quantities for the higher-dimensional dS spacetime with the non-linear charge correction. Note that the volume of this system is the geometrical volume between the dS black hole horizon and cosmological horizon and its conjugate thermodynamic quantity is the effective pressure, instead of the cosmological constant. Then we analyzed the thermodynamic property of phase transition for the four-dimensional dS spacetime with different non-linear charge corrections. We found the critical radio, horizon (of the dS black hole), and effective pressure are all increasing monotonically with the charge correction $\bar\phi$, while the critical radius (for the cosmological horizon), volume, effective temperature, and entropy are decreasing monotonically. Especially, with the increasing of the non-linear charge correction the two horizons get closer and closer, and the correction entropy $\bar f(x)$ is negative and indicates the interaction between the two horizons stronger and stronger.

In order to obtain the phase transition point, one can use the Maxwell's equal-area law or the Gibbs free energy. Here we point out that the results obtained by these two methods only coincide if the thermodynamic first law holds. For the low effective temperature, we found there exists the phase transition in $P_{eff}-V$ with different non-linear charge corrections. From the view of the Gibbs free energy, we also found the classical swallow tail behavior in $G_{P_{eff}}-T_{eff}$. The phase transition points in $P_{eff}-V$ and $G_{T_{eff}}-P_{eff}$ are the same for the given effective temperature $T_{eff}<T_{eff}^c$. And there were schottky peaks in $C_{P_{eff}}-T_{eff},~\alpha_{P_{eff}}-x,~\kappa_{T_{eff}}-x$ nearby the critical point. These phase transition properties are similar to that in AdS black holes. It was unique that the capacity at constant volume is nonzero, which is completely contrary to that in AdS black holes. We can regard it as a difference between dS spacetime and AdS black hole.

Then we investigated the dynamic process of phase transition for this system. From the inspect of Gibbs free energy, we found there is the double-well in $G-r_+$. The two local minimums correspond to the stable large and small dS black holes and have the same depth. The local maximum stands for the unstable intermediate dS black hole and acts as a barrier between the stable large and small dS black holes. It was a signal of emerging phase transition from the point view of $G$. Next we studied the dynamical process of phase transition governed by the Forkker-Planck equation. By imposing the reflection boundary conditions and considering a Gaussian wave packet as the initial condition, we obtained the numerical result of the Forkker-Planck equation: the initial Gaussian wave packet at the stable large or small dS black holes decreases with increasing with time, however at the same time the other peak of $\rho(t,r_+)$ at the stable small or large dS black holes increases from zero to a same constant. That indicates that with increasing time the system will leave from the initial state to another state, until it becomes a coexistent state, which is consistent with the fact that the depth of two wells of $G$ are the same value.

Finally we considered the first passage time. By imposing the absorption boundary condition on the intermediate-potential black hole state and considering a Gaussian wave packet at the stable large and small dS black holes as the initial condition, we also obtained the numerical result of the Forkker-Planck equation: $\Sigma$ decays very fast in a short time and it drops faster with increasing effective temperature. And the behavior of $F_p$ is the similar for different effective temperatures. There exists a single peak near $t=0$ in $F_p$. This can be understood as a large number of first passage events occur in a short interval of time, and the probability distribution decays exponentially with time. From the effect of effective temperature on $F_p$, $\Sigma$, and $G$, we found the higher temperature, the faster probability decreases, the easier phase transition occurs, and the lower depth of the barrier is; otherwise the lower temperature, the slower probability decreases, the harder phase transition occurs, and the higher depth of the barrier is.

\section*{Acknowledgments}
We would like to thank Prof. Zong-Hong Zhu, Ren Zhao, and Meng-Sen Ma for their indispensable discussions and comments. This work was supported by the National Natural Science Foundation of China (Grant No. 11705106, 11475108, 11605107, 12075143), the Natural Science Foundation of Shanxi Province, China (Grant No.201901D111315), the Natural Science Foundation for Young Scientists of Shanxi Province, China (Grant No.201901D211441), the Scientific Innovation Foundation of the Higher Education Institutions of Shanxi Province (Grant Nos. 2020L0471, 2020L0472), and the Science Technology Plan Project of Datong City, China (Grant Nos. 2020153).

\end{document}